\documentclass[twocolumn,floats,floatfix,superscriptaddress,aps,pra]{revtex4}
\usepackage{amsfonts}
\usepackage{amssymb}
\usepackage{amsmath}
\usepackage{calc}
\usepackage{graphicx}
\usepackage{bm}
\usepackage{makecell}
\usepackage{booktabs}

\def\be{ \begin{equation} }
\def\ee{ \end{equation} }
\def\bea{ \begin{eqnarray} }
\def\eea{ \end{eqnarray} }
\def\bse{ \begin{subequations} }
	\def\ese{ \end{subequations} }
\def\ba{ \begin{array} }
	\def\ea{ \end{array} }
\def\bt{\begin{tabular}}
	\def\et{\end{tabular}}

\def\i{\,\text{i}}
\def\e{\,\text{e}}
\def\i{i}
\def\e{e}

\def\fromto{\leftrightarrow}

\def\U{\mathbf{U}}

\def\i{{\rm{i}}}

\def\be{ \begin{equation} }
\def\ee{ \end{equation} }
\def\bea{ \begin{eqnarray} }
\def\eea{ \end{eqnarray} }
\def\bse{ \begin{subequations} }
	\def\ese{ \end{subequations} }

\def\U{\mathbf{U}}

\def\i{i}
\def\e{e}

\def\fromto{\leftrightarrow}

\newcommand{\ket}[1]{\vert #1\rangle}

\begin{document}
	
	\author{Boyan T. Torosov}
	\affiliation{Institute of Solid State Physics, Bulgarian Academy of Sciences, 72 Tsarigradsko chauss\'{e}e, 1784 Sofia, Bulgaria}
	\author{Michael Drewsen}
	\affiliation{Department of Physics and Astronomy, Aarhus University, DK-8000 Aarhus C, Denmark}
	\author{Nikolay V. Vitanov}
	\affiliation{Department of Physics, St Kliment Ohridski University of Sofia, 5 James Bourchier blvd, 1164 Sofia, Bulgaria}

	\title{Efficient and robust chiral resolution by composite pulses}
	
	\date{\today}
	
	\begin{abstract}
		We introduce a method for detection of chiral molecules using sequences of three pulses driving a closed-loop three-state quantum system.
		The left- and right-handed enantiomers have identical optical properties (transition frequencies and transition dipole moments) with the only difference being the sign of one of the couplings.
		We identify twelve different sequences of resonant pulses for which chiral resolution with perfect contrast occurs.
		In all of them the first and third pulses are $\pi/2$-pulses and the middle pulse is a $\pi$-pulse.
		In addition, one of the three pulses must have a phase shift of $\pi/2$ with respect to the other two.
		The simplicity of the proposed chiral resolution technique allows for straightforward extensions to more efficient and more robust implementations by replacing the single $\pi/2$ and $\pi$-pulses by composite pulses.
		We present specific examples of chiral resolution by composite pulses which compensate errors in the pulse areas and the detuning of the driving fields.
		%
	\end{abstract}
	
	\maketitle
	
	\section{Introduction\label{Sec:intro}}
	
	Chirality plays a crucial role in many branches of contemporary science. Chiral molecules are, e.g., extremely important in chemistry, biotechnologies, and pharmaceutics. Although the two enantiomers of a particular chiral molecule have the same configuration with respect to a spatial reflection, their chemical properties may be completely different in specific environments. This is particularly important in pharmaceutics, where the chirality of a particular substance may be essential for the effect and efficiency of a drug. For instance, L-Carnitine is a widely used food supplement, which has some beneficial effects on the human body, while D-Carnitine is known to be toxic \cite{Carnitine}.
	
	The traditional methods for enantiomer detection and separation are based on complicated and expensive chemical techniques, such as crystallization, derivatization, kinetic resolution, and chiral chromatography \cite{Ahuja}.
	Alternatively, one may use chiroptical spectroscopy to break the symmetry of the enantiomers by interaction with circularly polarized light \cite{Chiroptical}.
	Some prevalent chiroptical methods are optical rotary dispersion \cite{Chiroptical}, circular dichroism \cite{dichroism}, vibrational circular dichroism \cite{Nafie,NafieStephens}, and Raman optical activity \cite{Nafie,Barron}.
	These methods rely on the magnetic-dipole interaction between the circularly polarized light and the molecules.
	Furthermore, methods based on linearly polarized light have been developed, using the much stronger electric-dipole interaction \cite{LinearLight}.
	These methods make use of the sign difference of some of the transition dipole moments of the two enantiomers, which is then mapped onto population differences by using quantum systems with three or four states driven in closed-loop interaction schemes, thereby creating interferometric linkages.
	This approach has been further developed by Shapiro and co-workers \cite{Shapiro}, who used concepts from adiabatic passage methods \cite{STIRAP} to detect and separate enantiomers.
	Following these ideas, a simpler and faster approach, based on shortcuts to adiabaticity \cite{STA}, has been recently proposed \cite{VitanovDrewsen}.
	Finally, rotational spectroscopy has been used to develop methods, such as microwave three-wave mixing (M3WM) \cite{M3WM, Li2008} for chiral analysis in gas-phase samples.
	
	
	In this work, we develop an alternative to the approach introduced in Ref.~\cite{VitanovDrewsen}.
	Instead of using partly overlapping delayed pulses, we present a systematic method for producing chirality-dependent population transfer, based on sequential resonant pulses.
	Because at each instant of time only one transition is driven, this approach allows for a straightforward generalization by substituting the single pulses with composite ones, therefore making the technique highly robust and accurate.
	In contrast to the approach in Ref.~\cite{VitanovDrewsen}, which requires precise control of the shortcut pulse, the present method is easier to implement and control, as it systematically separates the interaction on different transitions.
	In such a way we focus on each transition and optimize its driving by a suitable composite pulse and hence improve the overall performance.
	
	To this end, we model the chiral molecules as a delta-type system, where one of the three couplings, marked as $P,S,Q$, differs in sign in the two enantiomers (see Fig.~\ref{fig:scheme}).
	In Sec.~\ref{sec:ThreeSinglePulses} we introduce our method and show how we can map this sign difference into population transfer to different states, by using a sequence of resonant pulses with total area of just $2\pi$.
	In Sec.~\ref{Sec:CP} we demonstrate how composite pulses can be used to improve the accuracy and robustness of the method.
	In Sec.~\ref{Sec:conclusions} we summarize the conclusions.


	\section{Three single pulses\label{sec:ThreeSinglePulses}}
	
	\subsection{Description of the method}\label{Sec:method}
	
	\begin{figure}[tb]
		\includegraphics[width=0.9\columnwidth]{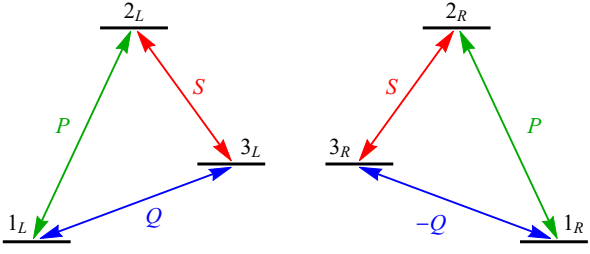}
		\caption{
			Coupling schemes for molecules with left ($L$) and right ($R$) handedness.
			The transition frequencies and the couplings are identical, the only difference being the sign of the transition dipole moment of the $Q$-transition, which results in Rabi frequencies of opposite signs.
		}
		\label{fig:scheme}
	\end{figure}

	We first describe the method to produce chirality dependent population transfer in a closed $\Delta$-type system, as illustrated in Fig.~\ref{fig:scheme}, by using suitable sequences of resonant pulses.
	As shown in Fig.~\ref{fig:scheme}, we assume that the only difference between the $L$- and $R$-handed molecules is in the sign of the $Q$-coupling driving the transition $\ket{1}\leftrightarrow \ket{3}$, and hence we neglect the (possible) extremely small differences due to the electroweak interaction \cite{Quack1998}.
	
	To be specific, let us consider the following general sequence of three resonant pulses: a $P$ pulse, followed by an $S$ pulse, followed by a $Q$ pulse.
	We can describe the evolution of the system by using the total propagator, which is a product of the three constituent propagators,
	\be\label{U-total}
	\U =\U_{Q} \U_S \U_P,
	\ee
	with
	\bse\label{UpUsUq}
	\begin{align}
	\U_P &= \left[ \begin{array}{ccc} a_P & b_P & 0 \\ -b_P^{\ast} & a_P^{\ast} & 0 \\ 0 & 0 & 1 \end{array}\right], \\
	\U_S &= \left[ \begin{array}{ccc} 1 & 0 & 0 \\0 & a_S & b_S  \\ 0 & -b_S^{\ast} & a_S^{\ast}  \end{array}\right], \\
	\U_Q &= \left[ \begin{array}{ccc}   a_Q & 0 & \pm b_Q  \\ 0 & 1 & 0 \\  \mp b_Q^{\ast} & 0 & a_Q^{\ast}  \end{array}\right],
	\end{align}
	\ese
	where the $\pm$ and $\mp$ signs in the last propagator account for $L$ (upper sign) and $R$ (lower sign) handedness.
	Here $a_X$ and $b_X$ $(X={P,S,Q})$ are the Cayley-Klein parameters of the three propagators $(|a_X|^2+|b_X|^2=1)$.
	For exact resonance and real-valued Hamiltonian, $a_X = \cos(A_X/2)$ and $b_X = -i \sin(A_X/2)$, where $A_X$ is the temporal pulse area.
	The total propagator \eqref{U-total}
	reads
	\be\label{totalProp}
	\U = \left[ \begin{array}{ccc}   a_P a_Q \pm b_P^{\ast} b_Q b_S^{\ast} &  a_Q b_P \mp a_P^{\ast} b_Q b_S^{\ast} & \pm a_S^{\ast} b_Q  \\
		-a_S b_P^{\ast} & a_P^{\ast} a_S & b_S \\
		a_Q^{\ast} b_P^{\ast} b_S^{\ast} \mp a_P b_Q^{\ast} & - a_P^{\ast} a_Q^{\ast} b_S^{\ast} \mp b_P b_Q^{\ast} & a_Q^{\ast} a_S^{\ast}  \end{array}\right].
	\ee
	
	Assume now that initially the system is in state $\ket{1}$, and the goal is to keep one enantiomer in this state at the end of the sequence, while the other is transferred to the state $\ket{3}$. In this case, only the first column of this matrix is important, and clearly the second entrance of this column should be zero, i.e., $U_{21}=0$.
	This means we must have $a_S=0$ or $b_P=0$, however, the latter case will, according to Eq. (3), clearly not be able to create a situation consistant with out goal of full chiral selectivity in population transfer.
	Hence the situation is now that we have $U_{11} = a_P a_Q \pm b_P^{\ast} b_Q b_S^{\ast}$, $U_{21} = 0$, and $U_{33} = \mp a_P b_Q^{\ast} + a_Q^{\ast} b_P^{\ast} b_S^{\ast}$.
	For \emph{in-phase} resonant excitations, the matrices in Eqs.~\eqref{UpUsUq} can always be brought on a form where all the $a_X$'s are real and all $b_X$'s are purely imaginary.
	In this case maximum chiral resolution cannot occur since the term $a_P a_Q$ in $U_{11}$ is always real while the terms $\pm b_P^{\ast} b_Q b_S^{\ast} $ will be imaginary.
	The solution of this problem is to add a phase shift of $\pi/2$ to one of the coupling fields, which will make one of the parameters $b_X$ real, such also the terms $\pm b_P^{\ast} b_Q b_S^{\ast}$ will be real, and our goal of maximum chiral selectivity is potentially possible.
	
	Let us now assume that the phase $\pi/2$ is applied to the coupling $Q$.
	Then a solution, which leads to a maximum contrast between $L$ and $R$,  is $a_P=a_Q=1/\sqrt{2}$ and $b_P=-i b_Q=-i/\sqrt{2}$, and $b_S=-i$.
	This is obtained if the $P$-pulse is a resonant $\pi/2$-pulse, the $S$-pulse is a resonant $\pi$-pulse, and the $Q$-pulse is a resonant $\pi/2$-pulse.
	Then we obtain the total propagators
	\be
	\U^{(L)} = \left[ \begin{array}{ccc} 0 & -i & 0 \\0 & 0 & -i  \\ -1 & 0 & 0  \end{array}\right],\quad
	\U^{(R)} = \left[ \begin{array}{ccc} 1 & 0 & 0 \\0 & 0 & -i  \\ 0 & -i & 0  \end{array}\right].
	\ee
	Hence, if the system starts in state $\ket{1}$, it will either stay in state $\ket{1}$ for the $R$ handedness or end up in state $\ket{3}$ for the $L$ handedness, as was our goal.

	\begin{table}
		\begin{tabular}{c c c}
			\hline
			\addlinespace[1ex]
			pulse sequence & final state ($L$) & final state ($R$)  \\
			\addlinespace[1ex]
			\hline
			\addlinespace[1ex]
			$P(\frac{\pi}{2}) iQ(\pi) S(\frac{\pi}{2})$ & 3 & 2  \\
			\addlinespace[1ex]
			$P(\frac{\pi}{2}) S(\pi) iQ(\frac{\pi}{2})$ & 3 & 1 \\
			\addlinespace[1ex]
			$iQ(\frac{\pi}{2}) P(\pi) S(\frac{\pi}{2})$ & 3 & 2  \\
			\addlinespace[1ex]
			$iQ(\frac{\pi}{2}) S(\pi) P(\frac{\pi}{2})$ & 1 & 2 \\
			\addlinespace[1ex]
			\hline
			\addlinespace[1ex]
			$P(\frac{\pi}{2}) Q(\pi) iS(\frac{\pi}{2})$ & 2 & 3  \\
			\addlinespace[1ex]
			$P(\frac{\pi}{2}) iS(\pi) Q(\frac{\pi}{2})$ & 1 & 3 \\
			\addlinespace[1ex]
			$Q(\frac{\pi}{2}) P(\pi) iS(\frac{\pi}{2})$ & 2 & 3  \\
			\addlinespace[1ex]
			$Q(\frac{\pi}{2}) iS(\pi) P(\frac{\pi}{2})$ & 2 & 1 \\
			\addlinespace[1ex]
			\hline
			
			\addlinespace[1ex]
			$iP(\frac{\pi}{2}) Q(\pi) S(\frac{\pi}{2})$ & 2 & 3  \\
			\addlinespace[1ex]
			$iP(\frac{\pi}{2}) S(\pi) Q(\frac{\pi}{2})$ & 1 & 3 \\
			\addlinespace[1ex]
			$Q(\frac{\pi}{2}) iP(\pi) S(\frac{\pi}{2})$ & 2 & 3  \\
			\addlinespace[1ex]
			$Q(\frac{\pi}{2}) S(\pi) iP(\frac{\pi}{2})$ & 2 & 1 \\
			\addlinespace[1ex]
			\hline
		\end{tabular}
		\caption{
			Pulse sequences and populated final state for the $L$- and $R$-handed enantiomers when the system starts in state 1 and is driven by three successive pulses.
			The numbers in the brackets correspond to the area of the pulse of the corresponding $P$, $S$ or $Q$ transitions, including a specific phase, if applicable.
		}
		\label{Table:sequences}
	\end{table}
	
\begin{figure}[tb]
	\includegraphics[width=8cm]{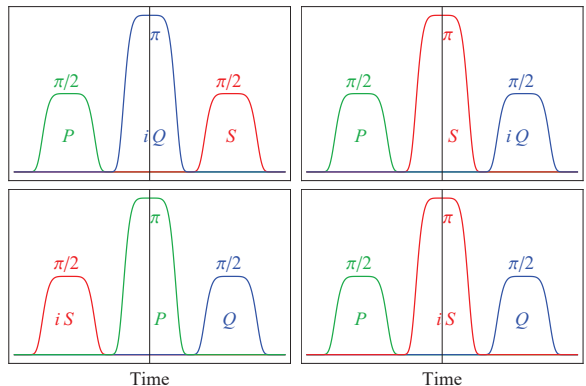}
	\caption{
		Examples of sequences of three resonant pulses that lead to chiral-dependent population transfer.
		It is supposed that the system is initially in state 1.
		The middle pulse has an area of $\pi$, and the other two pulses have areas of $\pi/2$ in all cases.	}
	\label{fig:SequenceThreeResonant}
\end{figure}

	This pulse sequence, which we denote as $P(\pi/2)S(\pi)iQ(\pi/2)$, is, however, not a unique solution to the chiral resolution problem.
	Other solutions exist where the final states of the two enantiomers differ.
	These are listed in Table \ref{Table:sequences}.
	All these twelve sequences have a similar structure: a $\pi$-pulse on one transition is sandwiched by $\pi/2$-pulses on the other two transitions. Four of these sequences are schematically illustrated in Fig.~\ref{fig:SequenceThreeResonant}, skipping  for brevity the other examples from Table~\ref{Table:sequences}.
	
	\begin{figure}[tb]
		\includegraphics[width=0.99\columnwidth]{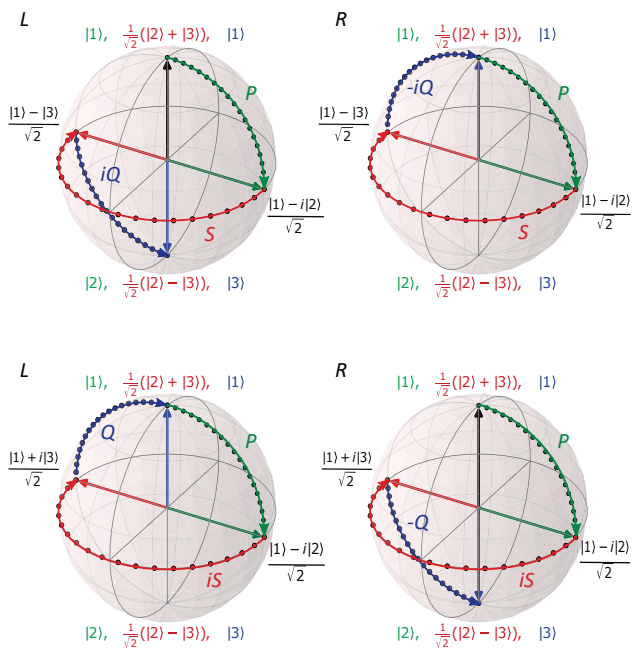}
		\caption{
			Evolution of our system, schematically illustrated on a pseudo Bloch sphere (See text for definition and interpretation), for the $P(\pi/2)S(\pi)iQ(\pi/2)$ sequence (top spheres) and for the $P(\pi/2)iS(\pi)Q(\pi/2)$ sequence (bottom spheres).
		}
		\label{fig:bloch}
	\end{figure}
	
	All cases which allow for chiral resolution have in common the fact that the first $\pi/2$-pulse acts either on the $P$ transition or the $Q$ transition.
	Either of these transitions connect the initially populated state 1 to another state, 2 or 3.
	In either cases, the $\pi/2$-pulse creates an equal coherent superposition of states 1 and 2, or 1 and 3.
	The following $\pi$-pulse transfers this coherent superposition to an equal coherent superposition on another transition.
	For instance, the coherent superposition of states 1 and 2 created by the $\pi/2$-pulse on the $P$ transition $1\fromto 2$ (top left frame of Fig.~\ref{fig:SequenceThreeResonant}) is transferred by the $\pi$-pulse on the $Q$ transition onto a coherent superposition of states 2 and 3.
	This superposition has a relative phase difference of $\pi$ for the $L$ and $R$ molecules because of the different sign of the $Q$ coupling.
	Hence, the last $\pi/2$-pulse on the $S$ transition beats with this coherent superposition and either puts the population to state 3 (for $L$) or 2 (for $R$): chiral resolution takes place.
	Similar interferometric processes, resulting in successful chiral resolution, occur in the other eleven cases in which the first $\pi/2$-pulse acts either on the $P$ transition or the $Q$ transition.
	
	As another example, in Fig.~\ref{fig:bloch} (top spheres) we illustrate the evolution of our system for the $P(\pi/2)S(\pi)iQ(\pi/2)$ sequence (top right frame of Fig.~\ref{fig:SequenceThreeResonant}) on a Bloch sphere. Starting from state $\ket{1}$ (north pole), the $\pi/2$ $P$ pulse transfers the system into the $(\ket{1}-i\ket{2})/\sqrt{2}$ state (green vector). After that the $\pi$ $S$ pulse creates another equal superposition, between the states $\ket{1}$ and $\ket{3}$ (red vector).
	Finally, the $\pi/2$ $i Q$ pulse drives the $L$ or $R$ enantiomers into the south or north pole of the Bloch sphere, respectively. The bottom spheres in the figure illustrate the evolution for the $P(\pi/2)iS(\pi)Q(\pi/2)$ sequence (bottom right frame of Fig.~\ref{fig:SequenceThreeResonant}).
	We note here that since we are dealing with a three-state system, each of these should not be considered as a usual Bloch sphere, but rather as a pseudo Bloch sphere representing each of the three operations with the north and south poles being the three different pair of states: $\ket{1}$ and $\ket{2}$, $(\ket{2}+\ket{3})/\sqrt{2}$ and $(\ket{2}-\ket{3})/\sqrt{2}$, and $\ket{1}$ and $\ket{3}$ for the $P$, $S$ and $Q$ transitions, respectively. These are shown as labels with green, red, and blue colors on the poles of the spheres in Fig.~\ref{fig:bloch}
	
	
	Until now we have mostly focused on the case where the phase shift of one of the fields are related to the coupling $Q$.
	As is clear from Table~\ref{Table:sequences}, we can assign it to any of the other two couplings, instead of $Q$.
	Let us assume that the phase shift $\pi/2$ is applied to $S$.
	Similar arguments, as for $iQ$, show that in the case of $iS$ we obtain chiral separation in the same four cases, but with reversed populations.
	The same final states as for $iS$ will be obtained if the phase shift $\pi/2$ is assigned to the coupling $P$.

	The availability of different pulse orders which lead to chiral resolution into different states allows one to select the most appropriate scheme for particular molecular structures and schemes for the consecutive state dependent separation process (e.g., by REMPI).
	For example, the $\pi$-pulse can be applied on the transition with the largest dipole moment.
	Then the $\pi/2$-pulses can be applied to the weaker transitions, thereby making the implementation easier.

	\section{Robust chiral resolution by composite pulses\label{Sec:CP}}
	
	\subsection{Compensation of pulse-area errors}
	
	Despite the benefits of its simplicity, in a practical situation the described approach may suffer from low fidelity, due to the reliance on resonant pulses of precise temporal area.
	We now show how one can improve the accuracy and robustness of this method by replacing the single pulses with composite pulse (CP) sequences:
	specifically, we replace the three single pulses with a composite $\pi/2$-pulse, followed by a composite $\pi$-pulse, followed by a composite $\pi/2$-pulse.
	In this sequence of three robust CPs, one expects that the phase of each CP must also play a role.
	To see how the global phases of the CPs affect the final state, let us suppose that the three composite propagators in the product presented in Eq.~\eqref{U-total} have the following form,
	\bse
	\begin{align}
	&\U_P=\left[ \begin{array}{ccc} \frac{1}{\sqrt{2}}\e^{\i\alpha_P} & \frac{1}{\sqrt{2}}\e^{\i\beta_P} & 0 \\ -\frac{1}{\sqrt{2}}\e^{-\i\beta_P} & \frac{1}{\sqrt{2}}\e^{-\i\alpha_P} & 0 \\ 0 & 0 & 1 \end{array}\right],
	\\
	&\U_S=\left[ \begin{array}{ccc} 1 & 0 & 0 \\0 &  0 &  \e^{\i\beta_S}  \\ 0 & -\e^{-\i\beta_S} &  0  \end{array}\right],
	\\
	&\U_Q=\left[ \begin{array}{ccc}   \frac{1}{\sqrt{2}}\e^{\i\alpha_Q} & 0 & \pm\frac{1}{\sqrt{2}}\e^{\i\beta_Q}  \\ 0 & 1 & 0 \\  \mp\frac{1}{\sqrt{2}}\e^{-\i\beta_Q} & 0 & \frac{1}{\sqrt{2}}\e^{-\i\alpha_Q}  \end{array}\right],
	\end{align}
	\ese
	where the $\pm$ sign stands again for $L$ or $R$ handedness and the phases $\alpha_X$ and $\beta_X$ have to be chosen appropriately.
	Starting from an initial state $\ket{i}=[1,0,0]^\text{T}$, it is easy to show that the sequence of these three propagators will produce the final state
	\be
	\ket{f}=\left[ \begin{array}{c} \frac{1}{2}\e^{\i(\alpha_P+\alpha_Q)}\pm \frac{1}{2}\e^{\i(\beta_Q - \beta_P - \beta_S)} \\
		0 \\
		\mp\frac{1}{2}\e^{\i(\alpha_P-\beta_Q)} + \frac{1}{2}\e^{-\i(\alpha_Q + \beta_P + \beta_S)} \end{array}\right].
	\ee
	From this expression, it is easy to see what condition needs to be fulfilled for the phases of the propagator, namely
	\be\label{phase-condition}
	\alpha_P + \alpha_Q + \beta_P + \beta_S = \beta_Q ,
	\ee
	or
	\be\label{phase-condition-1}
	\alpha_P + \alpha_Q + \beta_P + \beta_S = \beta_Q + \pi .
	\ee

	\begin{figure}[tb]
		\includegraphics[width=7cm]{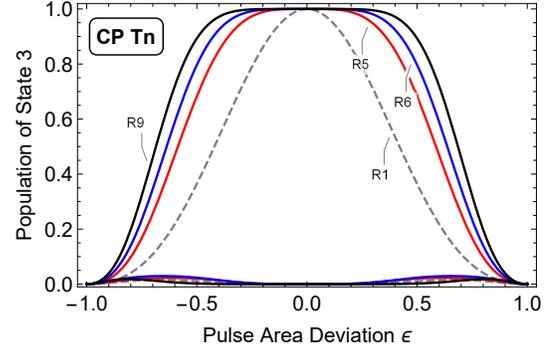}
		\caption{
			Population $P_3$ vs pulse area deviation for the sequence of three single resonant pulses $P(\frac{\pi}{2})iS(\pi) Q(\frac{\pi}{2})$ (dashed) and for a sequence of three CPs with pulse-area compensation \eqref{CPs}. The $Rn$ labels denote the $R$ curves for the corresponding $\text{T}n$ sequence. The $L$-curves lay close to zero and are almost indistinguishable.
		}
		\label{fig:vsArea}
	\end{figure}

	In the case of pulse-area-compensating CPs, it is straightforward to satisfy the condition \eqref{phase-condition} by
	setting
	\be\label{sym}
	\alpha_P = -\alpha_Q , \qquad
	\beta_P = \beta_Q , \qquad
	\beta_S = 0 .
	\ee
	The first two conditions are easily achieved by inverting the order of the constituent pulses in the $P$ and $Q$ CPs.
	Furthermore, an error in the pulse area would not break this symmetry condition and therefore no stabilization of the phase is needed in these two CPs.
	The third condition is trivial.
	However, setting the phase of the off-diagonal element of the propagator to zero is not enough, since an error in the pulse area will lead to deviation in the phase.
	Therefore, we need to use a CP with phase stabilization.
	To summarize, we need to use two half-$\pi$ CPs with inverted order of the constituent pulses, without phase stabilization (known as \emph{variable rotations}), and one $\pi$-pulse in the middle with phase stabilization (known as \emph{constant rotation}).
	The possibility to use variable rotations is very welcome because they can be realized by shorter composite sequences.
	
	Possible choices for composite $\pi/2$-pulses are \cite{ThetaPulses}
	\bse\label{CP-pi/2}
	\begin{align}
	C_X^{(1)}({\tfrac12\pi}) &= A_0 A_{\frac12\pi} , \\
	C_X^{(2)}({\tfrac12\pi}) &= A_0 B_{\frac23\pi} , \\
	C_X^{(3)}({\tfrac12\pi}) &= A_0 B_{\frac34\pi} A_{\pi} , \\
	C_X^{(4)}({\tfrac12\pi}) &= A_0 A_{\frac12\pi} A_0 A_{\frac32\pi} ,
	\end{align}
	\ese
	where $A=\pi(1+\epsilon)/2$ is a nominal half-$\pi$-pulse, $B=\pi(1+\epsilon)$ is a nominal $\pi$-pulse, and the subscripts at the r.h.s. denote the relative phases of the applied fields,
	while $X=P,S,Q$ denotes the transition on which the composite pulse is applied.
	We have introduced the dimensionless parameter $\epsilon$, which is used in this work as a measure of the deviation from the perfect value of the corresponding Rabi frequency (pulse area).
	
	Possible choices for phase-stabilized composite $\pi$-pulses are \cite{Wimperis1994}
	\bse\label{CP-pi}
	\begin{align}
	C_X^{(1)}(\pi) &= B_{\frac13\pi} B_{\frac53\pi} B_{\frac13\pi} , \\
	C_X^{(2)}(\pi) &=  B_{\chi} B_{3\chi} B_{3\chi} B_{\chi} B_{0} , \\
	\end{align}
	\ese
	where $\chi=\arccos(-1/4)$. 
	
	Figure \ref{fig:vsArea} shows the population of state 3 for left and right handedness for single pulses, $P(\frac{\pi}{2})iS(\pi) Q(\frac{\pi}{2})$, and for several sets of composite pulses, 
	\bse\label{CPs}
	\begin{align}
	&\text{T}5 = C_P^{(1)}({\tfrac12\pi}) C_{iS}^{(1)}(\pi) \overline{C}_Q^{(1)}({\tfrac12\pi}), \\
	&\text{T}6 = C_P^{(2)}({\tfrac12\pi}) C_{iS}^{(1)}(\pi) \overline{C}_Q^{(2)}({\tfrac12\pi}), \\
	&\text{T}7 = C_P^{(3)}({\tfrac12\pi}) C_{iS}^{(1)}(\pi) \overline{C}_Q^{(3)}({\tfrac12\pi}), \\
	&\text{T}9 = C_P^{(3)}({\tfrac12\pi}) C_{iS}^{(2)}(\pi) \overline{C}_Q^{(3)}({\tfrac12\pi}),
	\end{align}
	\ese
	where we denote by $\overline C^{(k)}_{X}(\frac12\pi)$ the same sequence as $C^{(k)}_{X}(\frac12\pi)$ but applied in the opposite order. For example, $\overline C^{(3)}_{X}(\frac12\pi) = A_{\pi} B_{\frac34\pi} A_0$.
	T5 is a sequence of 7 single pulses with a total pulse area of $5\pi$.
	T6 is a sequence of 7 single pulses with a total pulse area of $6\pi$.
	T7 is a sequence of 9 single pulses with a total pulse area of $7\pi$.
	T9 is a sequence of 11 single pulses with a total pulse area of $9\pi$.
	The notation T$n$ denotes the total pulse area $n\pi$ of each sequence, which is a measure of the total interaction duration.
	In Fig.~\ref{fig:vsArea} we compare the excitation profiles for the	$P(\frac{\pi}{2})iS(\pi) Q(\frac{\pi}{2})$ sequence with the corresponding CP analogues from Eq.~\eqref{CPs}. As seen from the figure, we obtain a much more robust excitation, and therefore distinction between $L$ and $R$ handedness using CPs, compared to the case of single pulses.

	\subsection{Compensation of pulse area and detuning errors}
	
	In a real situation, the energies of states 1, 2 and 3 might differ for the different molecules in the ensemble due to various sources of inhomogeneous broadening, due to spatially non-uniform magnetic field distributions and non-constant AC Stark shifts (light shifts) due to intensity variation of the applied fields.
	This would lead to detuning errors, which have to be compensated simultaneously with the pulse area error.
	
	The most straightforward way to address the problem of errors in both the pulse area and the detuning, might be to use a method similar to the one described in the previous subsection.
	However, the presence of detuning in $\text{CP}_P$  and $\text{CP}_Q$ would break the symmetry condition \eqref{sym}, if the phases are not stabilized.
	Hence, we need to use two $\pi/2$-CPs, and a $\pi$-CP in between, all with double compensation and stable phases in the composite propagators.
	
	To our knowledge, such pulses are not available in the literature.
	Therefore, we have derived numerically a few CPs with such properties.
	The five-pulse sequences that act as an error-compensated $\pi/2$-pulse, have the form \cite{TVtbb}
	\be
	D_X({\tfrac12\pi}) = A'_{\phi_1} B_{\phi_2} B_{\phi_3} B_{\phi_2} A'_{\phi_1},
	\ee
	where $A' = 0.4556\pi (1+\epsilon)$ and two options for the phases,
	\bse
	\begin{align}\label{CP5halfpiA}
	& \phi_1 = 0.5448\pi, \phi_2 = 0.3476\pi, \phi_3 = 0.9035\pi, \\ \label{CP5halfpiB}
	& \phi_1 = 0.5448\pi, \phi_2 = 1.2358\pi, \phi_3 = 0.6799\pi.
	\end{align}
	\ese
	These CPs compensate first-order errors in the overall propagator: $O(\epsilon)$ in the pulse area $\epsilon$ and $O(\delta)$ in the detuning $\delta$.
	
	Two five-pulse sequences that act as an error-compensated $\pi$-pulse, have the form \cite{TVtbb}
	\bse
	\begin{align}\label{CP5piA}
	D_X^{(1)}({\pi}) &= B_{\frac56\pi} B_{\frac23\pi} B_{\frac76\pi} B_{\frac23\pi} B_{\frac56\pi}, \\ \label{CP5piB}
	D_X^{(2)}({\pi}) &= B_{\frac56\pi} B_{\frac53\pi} B_{\frac76\pi} B_{\frac53\pi} B_{\frac56\pi}.
	\end{align}
	\ese
	They compensate simultaneous errors in the pulse area and the detuning up to the first order, $O(\epsilon)$ and $O(\delta)$.
	
	\begin{figure}[tb]
		\includegraphics[width=0.95\columnwidth]{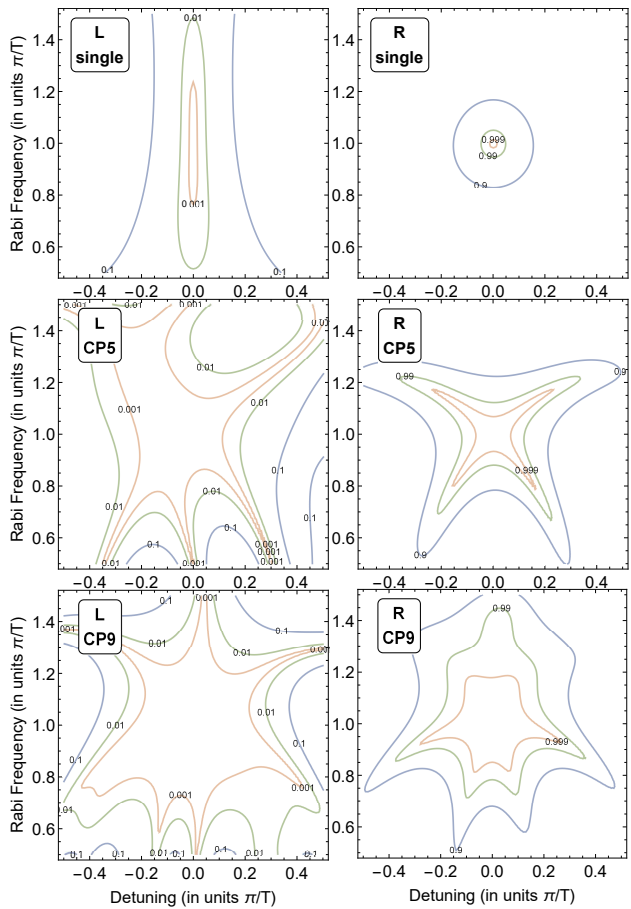}
		\caption{
			Population $P_3$ vs Rabi frequency and detuning for a single-pulse sequence (top) and for a sequence of three CPs with double compensation  of the form $D_P({\tfrac12\pi}) D_{iS}({\pi}) D_Q({\tfrac12\pi})$.
			The CP5 profiles (middle) are generated by the composite sequences from Eqs.~\eqref{CP5halfpiB} and \eqref{CP5piB}, and the CP9 profiles (bottom) are generated by the sequences of Eqs.~\eqref{CP9} and \eqref{CP9pi}. All pulses have rectangular pulse shape.
		}
		\label{fig:CpDouble}
	\end{figure}
	
	One may use longer sequences in order to achieve even better robustness of the excitation profile vs errors in the Rabi frequency and the detuning.
	For instance, we have derived the following 9-pulse sequence that acts as an error compensating $\pi/2$ pulse \cite{TVtbb},
	\bse\label{CP9}
	\be
	D_X({\tfrac12\pi}) = (A_1)_{\phi_1} (A_2)_{\phi_2} \cdots (A_5)_{\phi_5} \cdots (A_2)_{\phi_2} (A_1)_{\phi_1} 
	\ee
	where the nominal pulse areas [skipping the $(1+\epsilon)$ term for the sake of brevity] and the phases are
	\begin{align}\notag
	A_1 = 0.6771\pi, A_2 = 0.8579\pi, A_3 =0.6623\pi, \\
	A_4 = 0.5174\pi, A_5 =0.8812\pi,\label{CP9A} \\
	\phi_1 = 1.7517\pi, \phi_2 = 0.9043\pi, \phi_3 = 0.8820\pi, \notag\\
	\phi_4 = 0.9809\pi, \phi_5 = 1.6481\pi.\label{CP9F}
	\end{align}
	\ese
	Similarly, the following 9-pulse sequence \cite{TVtbb},
	\bse\label{CP9pi}
	\be
	D_X({\pi}) = B_{\phi_1} B_{\phi_2} B_{\phi_3}B_{\phi_4}B_{\phi_5} B_{\phi_4} B_{\phi_3} B_{\phi_2} B_{\phi_1},
	\ee
	where
	\begin{align}\notag
	\phi_1 = \pi/3, \phi_2 = 0.7379\pi, \phi_3 = 1.8092\pi, \\
	\phi_4 = 1.7379\pi, \phi_5 = 2\pi/3,
	\end{align}
	\ese
	acts as an error compensating $\pi$ pulse.
	The nine-pulse composite sequences compensate simultaneous errors in the pulse area and the detuning up to the second order, $O(\epsilon^2)$, $O(\delta^2)$ and $O(\epsilon\delta)$.

	All of the derived sequences lead to robust transition profiles vs both the Rabi frequency and the detuning and therefore provide robust separation of the chiral molecules in the ensemble.
	This can be seen from Fig.~\ref{fig:CpDouble}, where we compare the population of state $\ket{3}$ for the	$P(\frac{\pi}{2})iS(\pi) Q(\frac{\pi}{2})$ sequence with the corresponding CP analogues for the derived five-pulse (middle frames) and nine-pulse sequences (bottom frames).
	As seen from the figure, composite pulses may provide a better separation of chiral molecules in the presence of both pulse area and detuning error.
	Already the first-order error compensating five-pulse sequences provide a significant improvement over the single-pulse case.
	Further improvement is delivered by the second-order error compensating nine-pulse sequences.

	\section{Conclusions\label{Sec:conclusions}}
	
	In this work we have developed a general approach to detect chirality of molecules using sequences of three pulses in a closed three-state system. Our method relies on the opposite signs of one of the couplings in the system for the two enantiomers, which leads to constructive or destructive interference in the two enantiomers, and therefore to chiral resolution. The three interactions, used in our technique, are not overlapped, which allows us to treat them separately. In turn, this separation makes it possible to improve their accuracy and robustness by replacing the single pulses with composite pulses, and consequently, to optimize the overall performance of the method.

	\acknowledgments
	This work is supported by the European Commission's Horizon-2020 Flagship on Quantum Technologies project 820314 (MicroQC). MD acknowledges support from the Independent Research Fond Denmark, the European Commission's
	Horizon-2020 FET OPEN Project 766900 (TEQ) and the Villum Foundation.


\end{document}